# *DRUG REPURPOSING TO FIND INHIBITORS OF SARS-CoV-2 MAIN PROTEASE*

*Emilio Angelina[a) ], Sebastian Andujar[b)], Oscar Parravicini[b)], Daniel Enriz[b)] and Nelida Peruchena[a)]*

*[a)]Lab. Estructura Molecular y Propiedades, IQUIBA-NEA, Universidad Nacional del Nordeste, CONICET, FACENA, Av. Libertad 5470, Corrientes 3400, Argentina*
*[b)]Instituto Multidisciplinario de Investigaciones Biológicas (IMIBIO-SL), CONICET, San Luis, Argentina*

**ABSTRACT**

Severe acute respiratory syndrome coronavirus 2 (SARS-CoV-2) is the strain of coronavirus that causes coronavirus disease 2019 (COVID-19), the respiratory illness responsible for the COVID-19 pandemic. Currently there is no known vaccine or specific antiviral treatment for COVID-19 and so, there is an urgent need for expedite discovery of new therapeutics to combat the disease until a vaccine will be available worldwide.

Drug repurposing is a strategy for identifying new uses for approved drugs that has the advantage (over conventional approaches that attempt to develop a drug from scratch) that time frame of the overall process can be significantly reduced because of the few number of clinical trial required.

In this work, a virtual screening of FDA-approved drugs was performed for repositioning as potential inhibitors of the main protease $M^{pro}$ of SARS-CoV-2. As a result of this study, 12 drugs are proposed as candidates for inhibitors of the $M^{pro}$ enzyme. Some of the selected compounds are antiviral drugs that are already being tested in COVID-19 clinical trials (i.e. ribavirin) or are used to alleviate symptoms of the disease (i.e. codeine). Surprisingly, the most promising candidate is the naturally occurring broad spectrum antibiotic oxytetracycline. This compound has largely outperformed the remaining selected candidates along all filtering steps of our virtual screening protocol.

If the activity of any of these drugs is experimentally corroborated, they could be used directly in clinical trials without the need for pre-clinical testing or safety evaluation since they are already used as drugs for other diseases.

**INTRODUCTION**

Coronavirus disease 2019 (COVID-19) is the disease caused by the severe acute respiratory syndrome coronavirus 2 (SARS-CoV-2) strain.[1] It has named so because its RNA genome is about 82% identical to that of the SARS coronavirus (SARS-CoV).[2]

The World Health Organization (WHO) declared the 2019-20 coronavirus outbreak a Public Health Emergency of International Concern (PHEIC) on 30 January 2020 and a pandemy on 11 March 2020.[3] As of 15th May 2020 SARS-CoV-2 has caused more than four million of infections all around the globe and more than three hundred thousand deaths.

Currently, there is no vaccine or specific antiviral treatment for COVID-19.[4] Therefore, there is an urgent need, today more than ever before, for expedite discovery of novel therapeutics. Since the process of bringing a new pharmaceutical drug to the market, once a lead compound has been identified, usually takes from 10-15 years, a workaround to shorten the drug discovery cycle is required in the current scenario.

Drug repurposing or repositioning is a strategy for identifying new uses for approved drugs that are outside the scope of the original medical indication.[5] This strategy offers various advantages over developing an entirely new drug for a given indication. The most relevant one in the context of the current COVID-19 pandemy is the time frame for drug development that can be significantly reduced because most of the preclinical testing, safety assessment and formulation development already have been completed.

Moreover, since the onset of COVID-19 outbreak in Wuhan in December 2019, a lot of scientific resources related to SARS-CoV-2 have been released in expedited time, from the virus genome sequence[2] to protein structures solved by X-Ray crystallography to supramolecular structures mapped at atomic scale by CryoEM,[6, 7] which makes it possible to apply a Structure-based drug approach for drug discovery.

In the Protein Data Bank there are currently deposited both structural proteins of SARS-CoV-2 like the spike glycoprotein (S) and also non-structural proteins (NSPs) like the main protease ($M^{pro}$), papain-like protease ($PL^{pro}$), RNA-dependent RNA polymerase (RdRp) among other NSPs.

The S glycoprotein on the virion surface mediates receptor recognition and membrane fusion to gain entry into host cells. Biophysical and structural evidence reveal that both SARS-CoV and SARS-CoV-2 share the same functional host cell receptor, angiotensin-converting enzyme 2 (ACE2) but the novel strain binds to it with 10 to 20-fold higher affinity than the former which may contribute to the apparent ease with which SARS-CoV-2 can spread.[7] Because of its indispensable function, it represents a target for antibody-mediated neutralization. However, for small molecule drug design, protein-protein interaction (PPI) surfaces like that in S-ACE2 complex is very difficult to target because of its large size and flatness. Compared to the active site of an enzyme, PPIs are stabilized through large interfaces and it has been considered a difficult task to find small molecules competing with the binding of a protein partner. [8]

Regarding NSPs, RNA-dependent RNA polymerase (RdRp) is an enzyme required for SARS-CoV-2 replication and it was shown that can be inhibited by remdesivir, a broad spectrum antiviral drug that has been originally developed to treat Ebola virus disease.[9] On 29 April 2020, the National Institute of Allergy and Infectious Diseases (NIAID) announced that remdesivir was better than a placebo in reducing time to recovery for people hospitalized with advanced COVID-19 and lung involvement.[10] Although remdesivir has now been authorized for emergency use in the U.S., it is far from being the “silver bullet” to cure COVID-19.

Moreover, viral proteases are often druggable targets, HIV protease inhibitors are perhaps the most successful story about viral protease inhibitors development. Similarly to HIV protease, SARS-CoV-2 main protease ($M^{pro}$) is essential for processing the polyproteins that are translated from the viral RNA. Therefore, inhibiting the activity of this enzyme would eventually stop viral replication.[11]

One of the treatments options currently under study as part of the solidarity trial launched by the WHO[12] involves two HIV protease inhibitors, Lopinavir and Ritonavir. Regrettably, preliminary results conducted in adults hospitalized with severe COVID-19 has not shown benefit from this treatment.[13]

In this work we performed a Virtual Screening (VS) study for drug repurposing of FDA approved drugs as potential SARS-CoV-2 $M^{pro}$ inhibitors.

$M^{pro}$ is a dimer of two identical subunits (Figure 1), with each protomer composed of three domains. Two antiparallel β-barrel structures conform domains I and II and a third globular α-helical domain III is connected to domain II by means of a long loop region.[14] The substrate binding cleft is located between domains I and II.

The fold of the first two domains resemble architecture of serine proteases like chymotrypsin, but a cysteine amino acid and a nearby histidine (catalytic dyad) perform the protein-cutting reaction. The third α-helical domain is involved in regulating the dimerization of $M^{pro}$.[11]

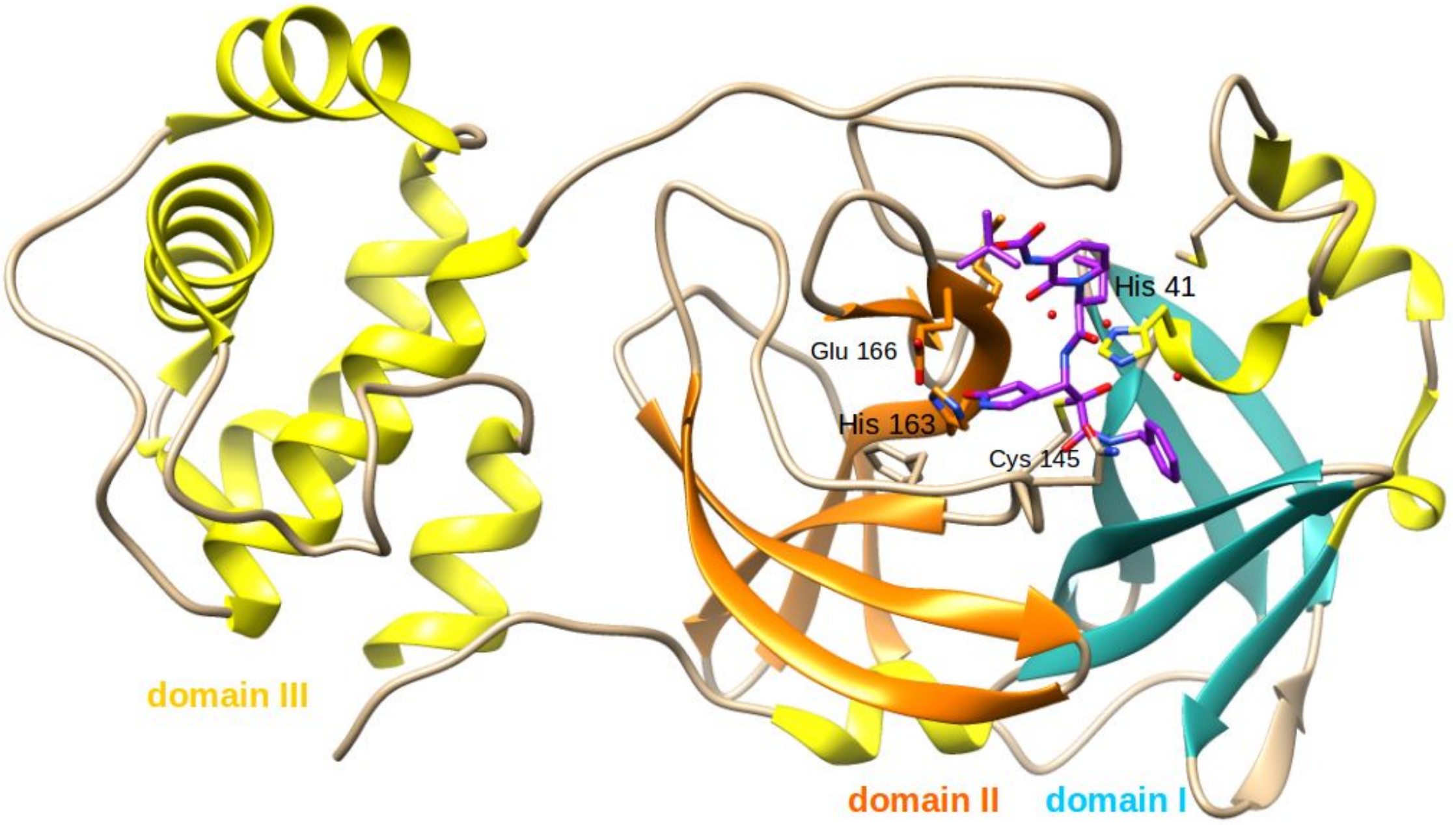

**Figure 1**. Structure $M^{pro}$ (monomer A) in complex with potent α-ketoamide inhibitor 13b from reference 11 (PDB code 6Y2G).

Hilgenfeld and colleagues have recently designed and synthesized peptidomimetic α-ketoamides to be used as inhibitors of the main proteases of coronaviruses. They obtained a potent inhibitor of the SARS-CoV-2 $M^{pro}$.[11] In another study, Jin and colleagues identified a mechanism-based inhibitor, N3, by computer-aided drug design and subsequently determined the crystal structure of SARS-CoV-2 $M^{pro}$ in complex with this compound.[14] Solved structures of the enzyme in complex with these two potent peptidomimetic inhibitors (PDB accession codes 6Y2G and 6LU7, respectively) provides structural basis for design of novel drug-like inhibitors. Moreover, another recent crystallographic fragment screening against SARS-CoV-2 $M^{pro}$ that has been deposited in the Protein Data Bank (DOI: 10.2210/pdb5rgj/pdb) also might help to find structural determinants for ligand anchoring within the enzyme binding cleft. In this work we exploited the available structural information about ligand binding to $M^{pro}$ binding cleft to guide the drug repurposing Virtual Screening experiment.

## COMPUTATIONAL DETAILS

Figure 2 shows the overall workflow for selection of candidate compounds.

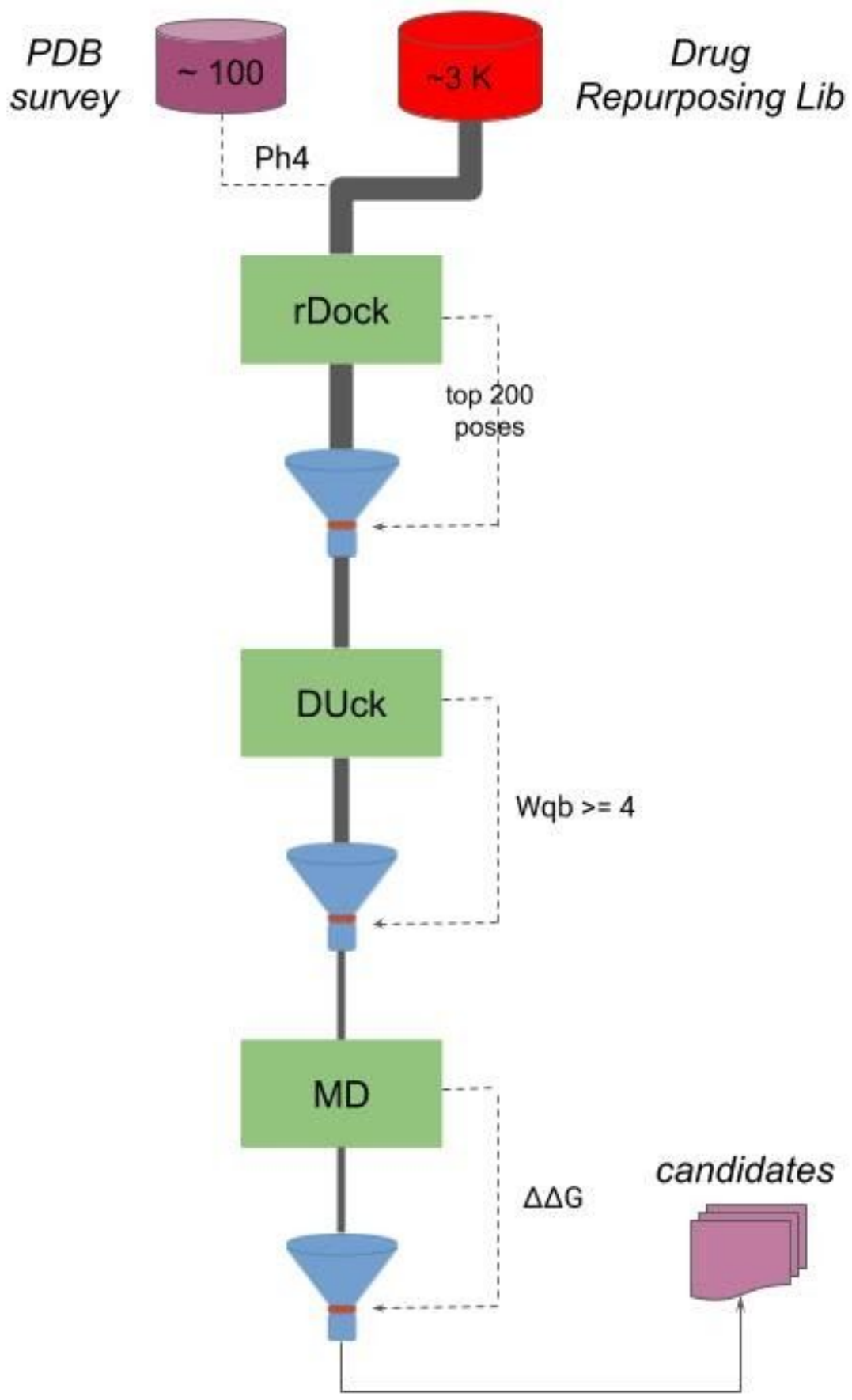

**Figure 2**. Workflow of the Virtual Screening procedure for candidate compound selection.

FDA-approved drugs collection compiled as 3D SDF files was downloaded from BindingDB database ([www.bindingdb.org](http://www.bindingdb.org)). LigPrep by Schrodinger was used to generate the different protomers and tautomers while retaining specified chiralities.[15] After generating the different chemical possibilities, the drug repurposing library ended up with ~ 3000 compounds.
Solved structure of unliganded form of SARS-CoV-2 $M^{pro}$ was employed as receptor for rigid docking (accession code: 6YB7). Up to date, more than 100 structures of SARS-CoV-2 $M^{pro}$ with ligands bound at the enzyme binding cleft, have been released. These ligand-bound structures of the enzyme were exploited in this work to find key interaction sites or "hot spots" within the binding cleft. These "hot spots" were then incorporated as pharmacophoric restraints (ph4s) to guide docking solutions.
Docking of drug repurposing library was performed with rDock program.[16] The cavity was prepared with the reference ligand method, using coordinates of inhibitor N3 covalently bound to SARS-CoV-2 $M^{pro}$(accession code 6LU7). N3 is a large peptidomimetic inhibitor that target all the binding cleft sub-pockets. Structure of 6LU7 was aligned to 6YB7 before performing the cavity mapping.
For each docked ligand, poses were sorted by total SCORE from standard rDock scoring function (SF) and only the top ranked pose was kept. Then, ligands best poses were sorted based in SF terms that account for intermolecular interactions (SCORE.INTER) and satisfaction of pharmacophoric restraints (SCORE.RESTR), whenever applicable. Top 200 ranked drugs (i.e. those that achieve the stronger interactions and satisfy the imposed restraints) were selected for DUck simulations (Figure 2).
DUck is a fast protocol for Steered Molecular Dynamic (SMD) simulations that has been thought for using in drug discovery projects. DUck calculates the work necessary to reach a quasi-bound state at which the ligand has just broken the most important native contact with the receptor ($W_{QB}$). Since true ligands form more resilient interactions than decoys, this non-equilibrium property has shown to be surprisingly effective in virtual screening contexts.[17]
To drive the DUck simulations we choose one of the "hot spots" identified in the previous PDB survey of available SARS-CoV-2 $M^{pro}$ complexes, i.e. the sames that were also considered for defining the ph4 restraints to guide docking calculations (see results section). Those ligands requiring a $W_{QB}$ > 4 kcal/mol to break the reference hydrogen bond with the enzyme were further scrutinized by MD simulations and binding free energy calculations. While DUck evaluates structural stability, binding free energy estimations provide information about thermodynamic stability of the complexes. Therefore, they are orthogonal to each other and can offer complementary information about complex stability. Accordingly, complexes that surpassed DUck filter were subjected to 20 nanoseconds of MD simulations in triplicate with Amber 16 software package (All-atoms force field ff14SB).[18] Those MD trajectory replicas from which ligand did not detached from enzyme binding cleft were sampled at regular intervals to get an averaged estimate of the relative binding free energy ($\Delta\Delta G$) of complexes by applying the MM-GBSA protocol. The details of this method have been presented elsewhere.[19] The conformational entropy change $-T\Delta S$ is usually computed by normal mode analysis, but in this study the entropy contributions were not calculated. In practice entropy contributions can be neglected when comparing states with similar entropy such as two ligands binding to the same protein, provided that only the relative free energy is informed.

**Quantum Theory of Atoms in Molecules (QTAIM)**
QTAIM analysis consists in the mapping of the gradient vector field $\nabla$ on top of the charge density (ρ) of the system thus giving rise to the topological elements of the charge density, two of which are unequivocal indicators of a bonding interaction: the Bond Critical Point (BCP) and the Bond Paths (BPs) that connect BCP to the bonded atoms.[20] Unlike the geometrical parameters (i.e. bond distance and angle and also types of atoms involved) for describing non covalent interactions, QTAIM does not rely on any arbitrary criteria for deciding whether an interaction is actually formed or not: the sole presence of a BCP and the corresponding BPs between two atoms guarantee the existence of the interaction.
The electronic charge density for complexes of SARS-CoV-2 $M^{pro}$ with selected candidate compounds was computed with Gaussian 16 software package.[21] The topological analysis of the charge density was then carried out with the help of Multiwfn[22] and *in house* python scripts. To sample a proper structure from the MD simulations for QTAIM analysis, trajectory frames were aligned by protein backbone atoms and then clusterized by similarity based on their Root Mean Square Deviations (RMSD). A representative structure from the most populated cluster was selected for QTAIM calculations. Since quantum mechanical calculations are still forbidden for full biomolecular complexes, reduced model systems were constructed from the selected representative structures by considering residues up to a distance range of 5 Å from ligand atoms.

**RESULTS AND DISCUSSION**

Figure 3 shows fragments from a recent reported crystallographic fragment screening on SARS-CoV-2 $M^{pro}$ (10.2210/pdb5rgj/pdb).

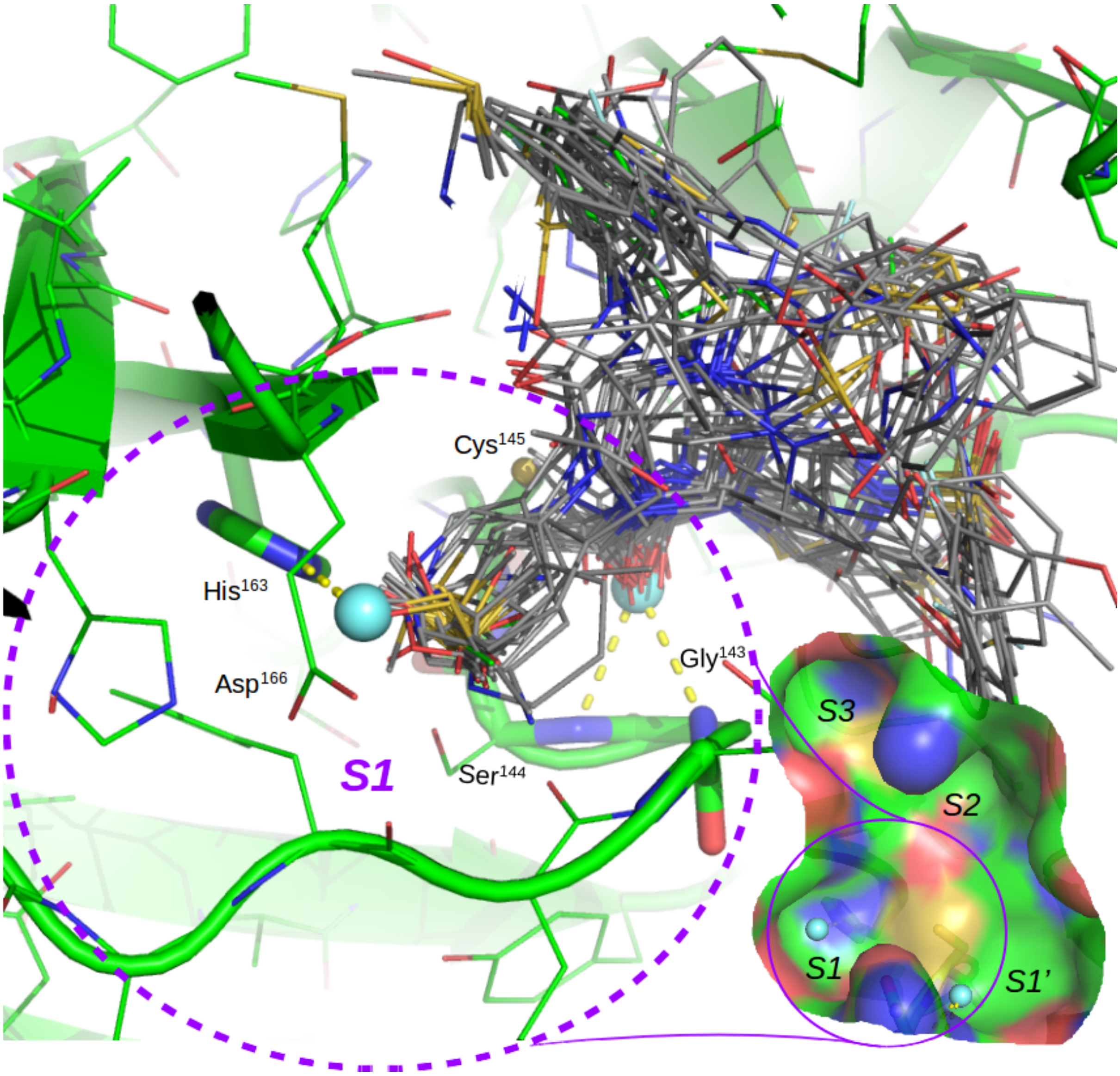

**Figure 3**. Fragments from a crystallographic fragment screening as they bound to $M^{pro}$ binding cleft. As observed in figure inset at the bottom right corner, the binding cleft is composed of several sub-pockets. At the S1 sub-pocket, we noted two conserved hydrogen bond (H-bond) sites among fragments that involves Gly 143 – Ser 144 backbone and His 163 imidazole. Pharmacophoric restraints (cyan spheres) at these interaction sites were considered to guide docking solutions in VS campaigns.

While different fragments bind to different regions of SARS-CoV-2 $M^{pro}$ binding cleft, there are two interaction sites at the enzyme S1 sub-pocket that are targeted by most fragments. They involve His 163 imidazole (site 1) and backbone of Gly 143 and Ser 144 residues from a loop structure that hold the catalytic Cys 145 (site 2). In both cases fragments are acting as H-bond acceptors against corresponding interaction site.

Interaction site 2 is specialized to hold an H-bond acceptor, since main-chain amides of Gly 143, Ser 144 and Cys 145 form the canonical “oxyanion hole” of the cysteine protease that stabilize the transition state negative charge on the amide carbonyl oxygen of the substrate after nucleophilic attack by Cys 145.[11]

Regarding interaction site 1, a closer look at His 163 neighborhood reveals that its interaction partner in most of the structures is not really one of the fragments from the crystallographic screen but it is actually a co-solvent molecule of DMSO. The fact that a DMSO molecule is almost constantly anchored at this site might indicate either that fragments do not have preference for it or that they are not able to displace the DMSO molecule. Solved structures of SARS-CoV-2 $M^{pro}$ in complex with peptidomimetic inhibitors (PDBs 6Y2G and 6LU7) show that inhibitor P1 $\gamma$-lactam moiety, designed as a glutamine surrogate, is deeply embedded in the S1 pocket of the protease where the lactam oxygen accept an H-bond from the imidazole of His 163.[11] Therefore, it is likely that His 163 interaction site is also important for ligand anchoring at the S1 sub-pocket.

As these interaction sites or “hot spots” seem to be important for ligand binding we have taken them into account in the virtual screenings in the way of pharmacophoric restraints, a functionality of the docking algorithm that is useful to incorporate knowledge-based features to guide docking solutions.

We first ran a virtual screening (VS) round without imposing pharmacophoric restraints and noted that most drugs tend to lay over the catalytic loop at the S1 sub-pocket forming H-bonds with backbone of Gly 143 and/or Ser144. Therefore, a restraint to force formation of these interactions might not be necessary in most cases. Nevertheless, to ensure anchoring on the catalytic loop, an H-bond acceptor restraint was set close to Gly 143 backbone (site). A gentle tolerance radius of 1.0 Å was asigned so that the ligand acceptor atom can satisfy the restraint by H-bonding to either Gly 143 N-H bond or any of the nearby equivalent bonds from the catalytic loop backbone (i.e. from Ser 144 or even Cys 145).

On the other hand, H-bond interaction with His 163 imidazole (site 1) was less frequently observed among docking poses and so, a restraint at this interaction site is mandatory to force H-bond formation. Accordingly, a second round of VS was run with pharmacophoric restraints at both interaction sites.

Top ranked compounds from both VS rounds were further scrutinized with methodology that allow a more accurate characterization of the binding event. Compounds that surpassed docking filters (see computational details section) were first subjected to DUck calculations. For compounds selected from the first unrestrained VS run, interaction site 2 was chosen to drive DUck calculations since H-bond to site 1 was rarely observed without imposing restraints, as discussed above. Whereas for compounds filtered from the second restrained VS round, both interaction sites were considered for $W_{QB}$ calculation and the highest value is informed (see Table 1 notes).

Standard MD simulations were then carried out in triplicate for compounds that performed well in DUck calculations in order to get an estimate of their relative binding energies ($\Delta\Delta G$) to $M^{pro}$ protease. For comparison purposes, binding energy was also computed for potent alpha-ketoamide inhibitor of SARS-CoV-2 $M^{pro}$ previously co-crystallized with the enzyme (accession code 6Y2G).

Table 1 reports candidate compounds selected from both VS rounds according to the protocol described in Figure 2. And figure 4 shows best docking poses of selected inhibitors from Table 1.

**Table 1**. Candidates compounds selected for experimental testing as non-covalent SARS-CoV-2 $M^{pro}$ inhibitors.

| Drug name | rDock Rank | $W_{QB}$ [a),b)] | Pose stability? [c),e)] | *ΔΔG* [a),d),e)] | VS run |
|---|---|---|---|---|---|
| Codeine (COD) | 7 | 4.01$^{(s1)}$ | yes (2/3) | 30.34 | restrained |
| Ticagrelor (TIC) | 11 | 4.44$^{(s2)}$ | NC | NC | restrained |
| Oxytetracycline (OTC) | 12 | 11.18$^{(s2)}$ | yes (3/3) | 12.39 | unrestrained |
| Ribavirin (RIBA) | 14 | 4.70$^{(s1)}$ | yes (2/3) | 39.25 | restrained |
| Chlortetracycline | 16 | 4.75$^{(s2)}$ | NC | NC | unrestrained |
| Losartan | 16 | 4.66$^{(s1)}$ | NC | NC | restrained |
| Doxycycline | 20 | 4.16$^{(s2)}$ | NC | NC | unrestrained |
| Dyphylline | 29 | 4.33$^{(s1)}$ | NC | NC | restrained |
| Regadenoson | 62 | 4.80$^{(s2)}$ | NC | NC | unrestrained |
| Etoposide | 108 | 4.40$^{(s2)}$ | NC | NC | unrestrained |
| Pentostatin (ECF) | 163 | 8.02$^{(s2)}$ | NC | NC | unrestrained |
| Elvitegravir (ELV) | 171 | 6.75$^{(s2)}$ | NC | NC | unrestrained |

a) In kcal/mol

b) $W_{QB}$ value necessary to break ligand interaction with site 1 or 2 as indicated by the labels (s1) or (s2).

c) A pose is considered to be stable if ligand remains attached to binding cleft during simulation in at least 1 of 3 replicas.

d) ΔΔG was computed with respect to potent $\alpha$-ketoamide inhibitor.

e) NC = Not Computed yet

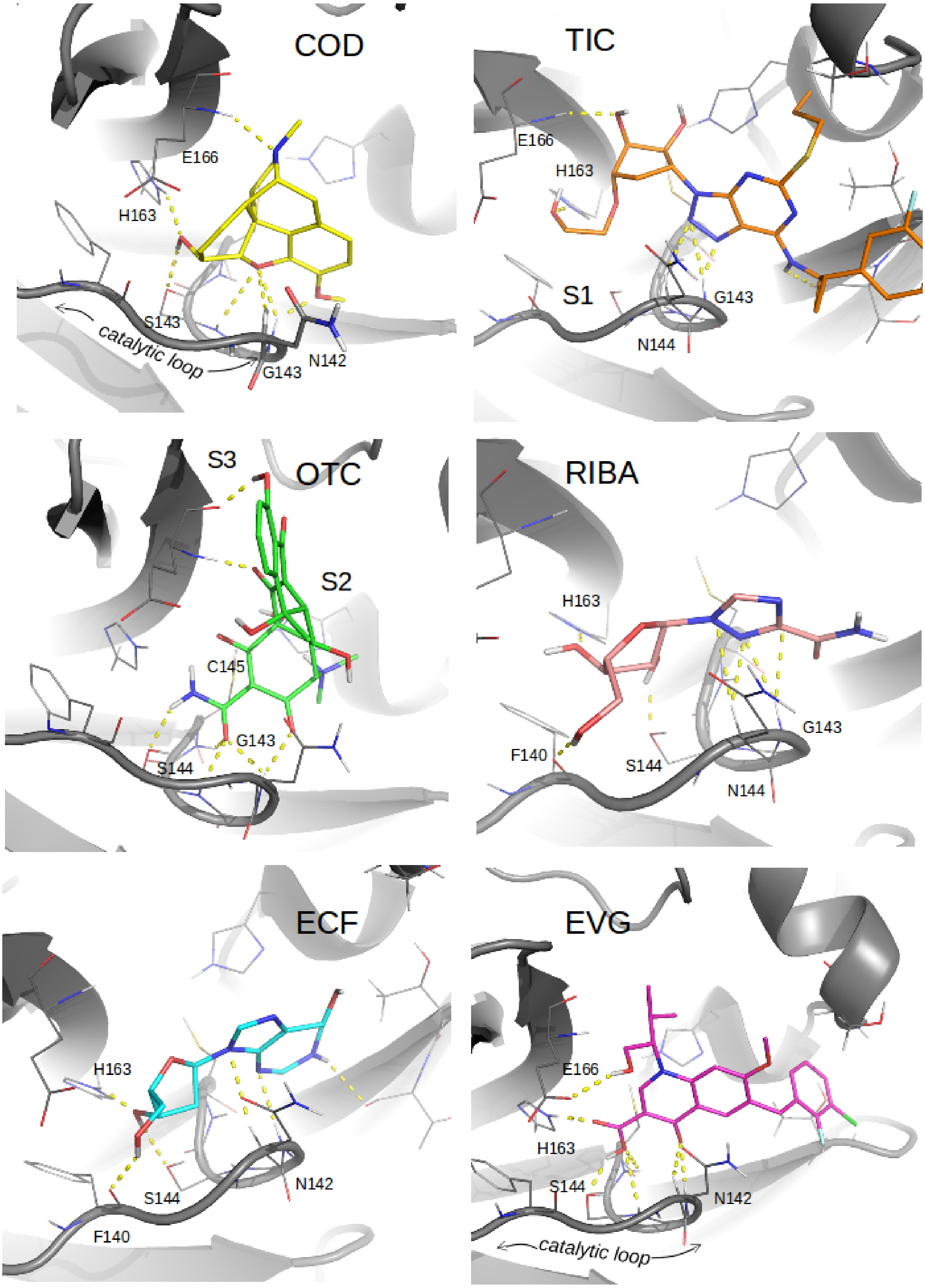

**Figure 4**. Docking poses for selected compounds from Table 1. COD = codeine, TIC = ticagrelor, OTC = oxytetracycline, RIBA = ribavirin, ECF = pentostatin, ELV = elvitegravir.

Compounds in Figure 3 can be considered as a representative sample of common scaffolds found among top ranked compounds from both VS rounds.
COD, ELV and OTC, despite having different overall structure they share a common pharmacophoric feature consisting in two acceptor oxygen atoms in a planar structure at 3-bonds distance to each other (i.e. in 1, 4 position). This acceptors disposition allows formation of several N–H···O interactions with H-bond donors from the catalytic loop backbone, at the S1 sub-pocket. At least one of the acceptor atoms is involved in a ring structure which is oriented nearly perpendicularly to the catalytic loop and flanked at both sides by side chain of Cys 145 and Asn 142.
On the other hand, nitrogen-rich heterocyclic rings from RIBA, TIC and ECF also tend to interact with the oxyanion hole donors but they are oriented parallel to the catalytic loop backbone forming N–H···$\pi$ interactions.
In addition to binding at the catalytic loop, additional anchoring to S1 sub-pocket is provided by interactions with side chain of His 163, Ser 144 and Glu 166 and backbone of Phe 140. Anchoring at this deepest region of S1 sub-pocket seems to be better achieved by compounds having a 5-membered sugar-like ring placed at this region (i.e. as in RIBA, ECF and TIC) that is able to form an extensive H-bond network through its OH groups.
Moreover, TIC, ECF and EVG also target S1' sub-pocket, the first two compounds form a strong N–H···O interaction with backbone carbonyl oxygen of residue Thr 26 from such sub-pocket.
Most compounds do not extend beyond S1 and S1' sub-pockets except OTC that also target S2 and S3 sub-pockets from enzyme binding cleft by forming two strong interactions with backbone of Glu 166, among other interactions (discussed below). COD and TIC partially extend into S2 sub-pocket by accepting a proton from backbone amide of Glu 166.
Regarding the known uses of these drugs, COD is already being used for clinical management of COVID-19 patients to suppress coughing when it is distressing. It could be the case that a therapeutic effect of COD (i.e. due to $M^{pro}$ inhibition) might be also operating under the hood in conjunction with its cough alleviating effect, to improve overall patient condition.
RIBA and ELV are antiviral medications whose known action mechanisms do not involve the viral protease. ECF similarly than RIBA is a purine analog that interfere with DNA synthesis and is used as anticancer treatment. TIC acts as a platelet aggregation inhibitor used for the prevention of stroke, heart attack and other events in people with acute coronary syndrome. Whereas OTC is a naturally occurring broad spectrum antibiotic produced as secondary metabolite by *Streptomyces* strains. Chlortetracycline and Doxycycline in Table 1 also belong to the same group of antibiotics.
As indicated by data in Table 1, OTC is the most promising candidate for repositioning as SARS-CoV-2 $M^{pro}$ inhibitor. As indicated by the $W_{QB}$ value, a work of at least ~11 kcal/mol is required to break main interactions with the oxyanion hole residues and to pull it out from the enzyme binding cleft. Moreover, as shown in Figure 5, its $W_{QB}$ profile as a function of H-bond distance resemble that of a strong ligand, i.e. with similar outcomes among all DUck replicas.

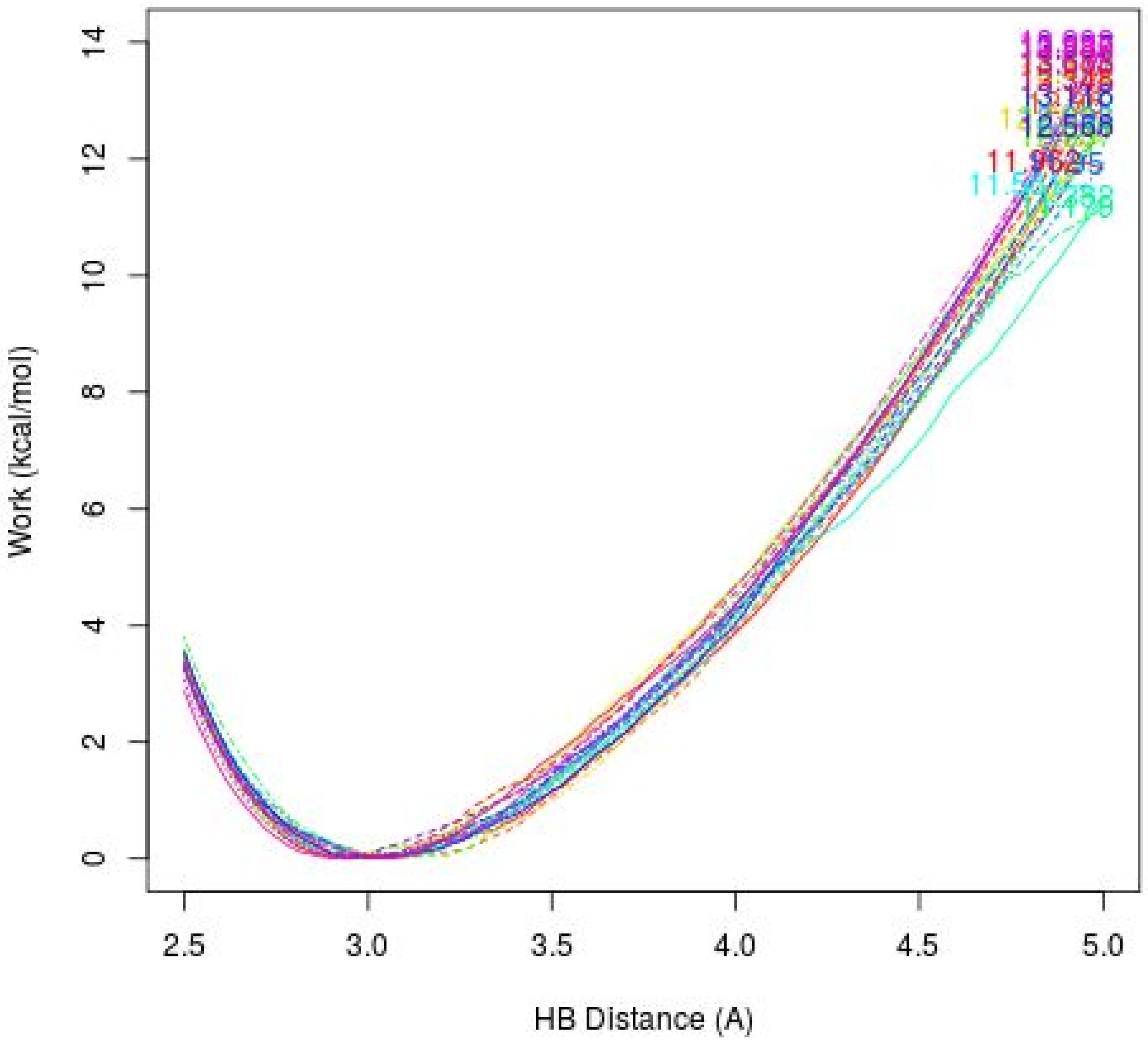

**Figure 5**. $W_{QB}$ profile for OTC.

OTC also preserves its original docking pose within M$^{pro}$ binding cleft along the entire simulation time in three out of three MD replicas (pose stability in Table 1). Stability of OTC binding is also reflected in the complex relative binding free energies (ΔΔG) which has been computed with respect to potent alpha-ketoamide inhibitor. For instance, binding of OTC to M$^{pro}$ is about 18 kcal/mol more stable than for COD and about 27 kcal/mol than for RIBA (see Table 1). Furthermore, binding of potent alpha-ketoamide inhibitor to M$^{pro}$ is just ~12 kcal/mol more stable than for OTC which is an encouraging result considering the larger size of the first one which is a large peptidomimetic inhibitor that target all sub-pocket of M$^{pro}$ binding cleft and it has been specifically designed to bind SARS-CoV-2 M$^{pro}$.[11]

**A more detailed analysis of OTC binding to M$^{pro}$**

For a more *in deep* analysis of the molecular interactions of OTC at the M$^{pro}$ binding cleft we next performed a topological analysis of the charge density (QTAIM) on a representative structure from MD simulations.

Figure 6 shows binding mode of OTC at the SARS-CoV-2 M$^{pro}$ binding cleft. Topological elements of the charge density describing OTC interactions are depicted.

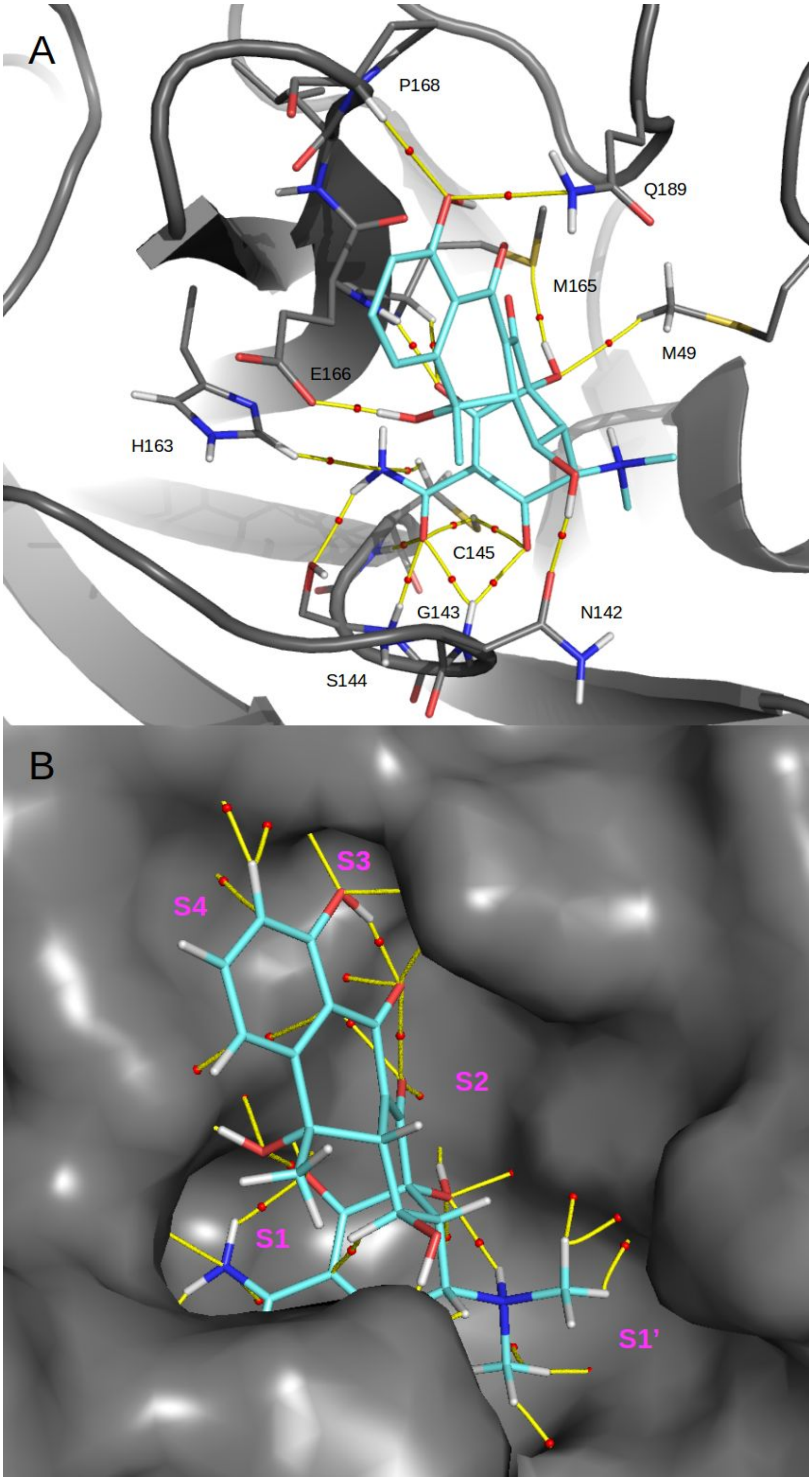

**Figure 6**. Binding mode of OTC at the SARS-CoV-2 $M^{pro}$ binding cleft. Topological elements of the charge density are depicted: small red circles are the Bond Critical Points (BCPs) and yellow lines connecting them to neighboring nuclei are the Bond Paths (BPs). In A only topological elements for the most relevant intermolecular interactions are depicted. While in B all the interactions involving OTC are shown to emphasize its extensive anchoring that encompass several binding cleft sub-pockets.

Perhaps the most salient characteristic of OTC as potential SARS-CoV-2 $M^{pro}$ inhibitor is its overall molecular shape which complements very well the enzyme binding cleft. $M^{pro}$ binding cleft is a J-shaped cavity where the S1 and S1' sub-pockets are perpendicularly oriented to S2 and S3 sub-pockets. Thanks to OTC curved shape and size, it can lay over S1 sub-pocket surface forming an intricate network of interactions with oxyanion hole residues Gly143, Ser144 and Cys145 (Figure 6A) and at the same time it can also bind to the S2 and S3 sub-pockets which provide additional anchoring points.

While the overall docking pose is preserved after MD simulations, there have been some changes in the interaction pattern that it is worth noting (compare with Figures 3). For instance, Glu 166 side chain which was too far as to directly interact with OTC in the docking pose, formed a strong H-bond with an OTC hydroxyl group after MD simulations. Also interactions at S2/S3 sub-pockets have slightly changed with respect to starting docking pose.

Regarding sub-pocket S2, it is composed by a flat surface as evidenced in Figure 6B. However, it has been shown that S2 features substantial plasticity, enabling it to adapt to the shape of inhibitor moieties.[11] Potent alpha-ketoamide inhibitor 13b (pdb code 6Y2G) bears a cyclopropyl methyl moiety at P2 that fits snugly into the enzyme S2 hydrophobic sub-pocket (see reference 11). Unlike 13b, OTC does not target this hydrophobic region of S2 which might explain the different shape that S2 adopts and also partially their differences in terms of binding affinity.

## CONCLUSIONS

In this work we have performed a virtual screening of FDA approved drugs for repurposing as potential inhibitors of SARS-CoV-2 main protease $M^{pro}$.

Although other studies for drug repositioning as inhibitors of SARS-CoV-2 molecular targets have already been reported, they focused mostly in antiviral medications. To our knowledge, oxytetracycline, which is the most promising candidate selected in the current virtual screening study has not been tested yet, neither in specific $M^{pro}$ inhibitory assays nor in viral replication assays. We encourage scientific community working on COVID-19 projects to include it (along with the remaining selected candidates) in their experimental screening pipelines.

## ACKNOWLEDGEMENT

We acknowledge SECyT-UNNE and CONICET for financial support. Geforce GTX Titan X GPU partially used for this research was donated by NVIDIA Corporation. The authors also thanks Prof Xavier Barril from University of Barcelona for his support and advices on writing this manuscript.